\documentstyle[12pt,a4]{article}

\newcommand{\mh}{m_H}
\newcommand{\gh}{\Gamma_H}
\newcommand{\ph}{\Pi_H}
\newcommand{\mw}{m_W}
\newcommand{\mz}{m_Z}
\newcommand{\mr}{m_R}
\newcommand{\WW}{W^+W^-}
\newcommand{\ZZ}{ZZ}
\newcommand{\VV}{VV}
\newcommand{\WZ}{W^\pm Z}
\newcommand{\ww}{W^\pm W^\pm}
\newcommand{\qq}{qq}
\newcommand{\GG}{gg}
\newcommand{\pvv}{\Pi_{VV}}
\newcommand{\A}{{\cal{A}}}
\newcommand{\B}{{\cal{B}}}
\newcommand{\Ab}{\bar{\A}}
\newcommand{\Ai}{\A_I}
\renewcommand\O{{\cal{O}}} % This replaces Danish capital Oslash
\newcommand{\aij}{a^I_J}
\newcommand{\ai}{a^I}
\renewcommand{\Im}{\mbox{Im}}
\newcommand{\beq}{\begin{equation}}
\newcommand{\eeq}{\end{equation}}
\newcommand{\beqn}{\begin{eqnarray}}
\newcommand{\eeqn}{\end{eqnarray}}
\newcommand{\beqns}{\begin{eqnarray*}}
\newcommand{\eeqns}{\end{eqnarray*}}

\catcode`\@=11
\newlength{\captionsize}
\newlength{\captionlength}
\long\def\@caption#1[#2]#3{\par\addcontentsline{\csname
  ext@#1\endcsname}{#1}{\protect\numberline{\csname
  the#1\endcsname}{\ignorespaces #2}}\begingroup
    \@parboxrestore
    \normalsize
    \@makecaption{\csname fnum@#1\endcsname}{\ignorespaces
\settowidth{\captionsize}{#1~\csname the#1\endcsname:}%
\setlength{\captionsize}{-\captionsize}%
\addtolength{\captionsize}{\textwidth}%
\addtolength{\captionsize}{-3mm}%
\settowidth{\captionlength}{#3}%
\ifdim\captionlength<\captionsize{#3}\else%
\parbox[t]{\captionsize}{#3}%
\fi%
}\par
  \endgroup}
\catcode`\@=12

\begin{document}

\begin{titlepage}
\noindent
\begin{flushright}
  CERN--TH/95--94\\
  hep-ph/9505211\\
  April, 1995
\end{flushright}
\vspace*{\fill}
\begin{center}
{\Huge\bf
  The Higgs Boson Lineshape\\
\ and Perturbative Unitarity
  }
\vspace*{2ex} \\
\large{\bf
  Michael H.~Seymour,} \\
  Division TH, CERN, \\
  CH-1211 Geneva 23, Switzerland.
\end{center}
\vspace*{4ex}

\subsection*{Abstract}
We discuss the lineshape of a heavy Higgs boson, and the behaviour well
above resonance.  Previous studies concluded that the energy-dependent
Higgs width should be used in the resonance region, but must not be used
well away from it.  We derive the full result and show that it smoothly
extrapolates these limits.  It is extremely simple, and would be
straightforward to implement in existing calculations.

\vspace*{\fill}
\vspace*{\fill}
\noindent
  CERN--TH/95--94\\
  April, 1995
\end{titlepage}

Gauge theories are carefully constructed so as to be well-behaved in the
high energy limit.  This means not only that they are finite, but that
they obey unitarity constraints.  This usually comes about by delicate
cancelations between different types of contribution, which means that
any small changes in the theory, for example anomalous couplings, show
up as very large changes in the high energy scattering amplitude.

In the particular case of the electroweak theory, it has long been
recognized that high energy scattering of $W$ and $Z$ bosons (which are
radiated from incoming quark lines in a hadron-hadron collision)
constitutes a crucial test of the theory[\ref{r1}].  Furthermore, if the
Higgs boson is heavy, $\mh\gg\mw,$ it will be directly seen as a
resonance in the $I\!=\!0$ scattering amplitude.  In this case,
effective theories have been derived that greatly simplify the
calculation of high energy scattering amplitudes[\ref{r2}], since the
vector bosons are replaced by the corresponding scalar Goldstone bosons.
We use the non-linear $\sigma$-model formulation[\ref{r3}], which has
advantages over the usual formulation for our purposes, because the
separation into resonant and non-resonant diagrams is the same as in the
electroweak theory, allowing a simpler interpretation of the final
result.  It should be stressed however that we use it purely as a
calculational device to reach this result, which is equally valid in the
full electroweak theory.

\begin{figure}[b]
  \vspace{4cm}
  \centerline{
    \includegraphics{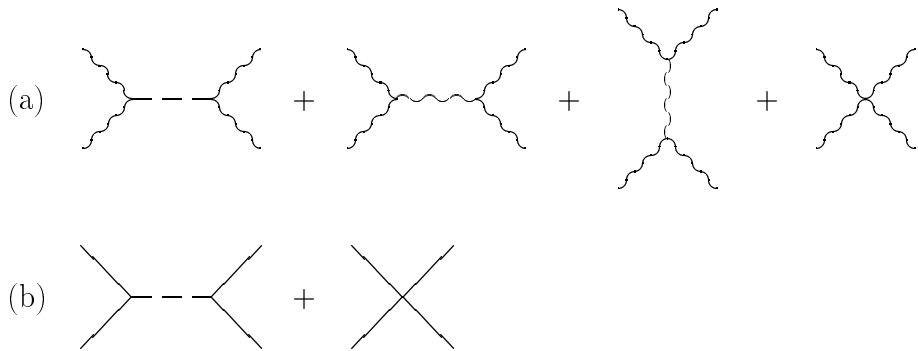}}
  \caption{Lowest order Feynman diagrams for $\WW\to\ZZ$ in the
    electroweak theory (a), and the
    high-energy effective theory~(b), in which the vector
    bosons are represented by the corresponding Goldstone bosons.}
  \label{f1}
\end{figure}

We begin by calculating the amplitude for $\WW\to\ZZ,$ from which all
others $\VV\to\VV$ can be derived by symmetry relations[\ref{r4}], which
we give later.  The lowest order Feynman diagrams are shown in
Fig.~\ref{f1}a, and again in the effective theory in Fig.~\ref{f1}b.
The effective theory correctly reproduces enhanced terms of order
$g^2\mh^2/\mw^2$ and $g^2s/\mw^2,$ but not the remainder of the order
$g^2$ amplitude (we assume $s,\mh^2\gg\mw^2,$ but make no assumption
about their relative size).  One gauge cancelation has already taken
place, since the last three diagrams of Fig.~\ref{f1}a are separately
$\sim s^2\mw^4$ but their sum, the second diagram of Fig.~\ref{f1}b,
is~$\sim s/\mw^2$.  The result for the amplitude is
\beq\label{e1}
  i\A = \frac{-ig^2}{4\mw^2} \left\{\frac{s^2}{s-\mh^2}-s\right\},
\eeq
where the two terms correspond to the two diagrams of Fig.~\ref{f1}b.
It is clear that at large $s\gg\mh^2,$ another cancelation occurs so
that the amplitude remains finite, and satisfies unitarity (except if
$\mh$ is very large),
\beq\label{e2}
  i\A \;\;{\stackrel{s\gg\mh^2}{\longrightarrow}}\;\;
\frac{-ig^2\mh^2}{4\mw^2},
\eeq
although the two contributions separately do not.

\begin{figure}[b]
  \vspace{1.9cm}
  \centerline{
    \includegraphics{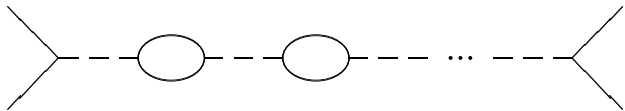}}
  \caption{All orders resummation leading to the Higgs boson
    self-energy.}
  \label{f2}
\end{figure}

In the resonance region, $s\sim\mh^2,$ it is clear that the amplitude
(\ref{e1}) diverges.  As is well known, this is regulated by resumming
to all orders the diagrams of Fig.~\ref{f2}.  Since each diagram
contains an additional factor of $1/(s-\mh^2),$ it is always the leading
diagram at that order, and the resummation is justified.  The result
is[\ref{r5}]
\beq\label{e3}
  i\A \;\;{\stackrel{s\sim\mh^2}{\longrightarrow}}\;\; \frac{-ig^2}{4\mw^2}
    \frac{s^2}{s-\mh^2+i\Im\ph(s)},
\eeq
where $\Im\ph(s)=\mh\gh s^2/\mh^4$ is the imaginary part of the Higgs
boson self-energy\footnote{Strictly speaking the $\mh^2$ in the
  denominator of (\ref{e3}) should have been replaced by a
  scheme-dependent parameter $\mr^2\sim\mh^2$[\ref{r3}], but for clarity
  we leave it as $\mh^2$ (or alternatively, choose a scheme in which
  they are identical)}.  Since $\gh/\mh\sim g^2\mh^2/\mw^2,$ including
the self-energy in the propagator promotes the resonant diagram by one
order at the resonance, so one is formally justified in neglecting the
non-resonant diagram.

However, inserting (\ref{e3}) into (\ref{e1}), one immediately sees that
the high energy behaviour is spoiled, since the resonant diagram is
suppressed and the cancelation no longer occurs.  The conventional
resolution is as follows:  {\em Outside the resonance region, the
  diagrams of Fig.~\ref{f2} are not enhanced, so there is no
  justification for resumming them.  Therefore the correct result is
  (\ref{e3}) in the resonance region and (\ref{e1}) outside it.}

While this statement is correct theoretically it is not very useful for
phenomenology, since one needs an amplitude that smoothly extrapolates
the different regions.  It is the main aim of this paper to calculate
such an amplitude.

Since we have stressed the importance of the cancelation between the
resonant and non-resonant diagrams well above resonance, it is natural
to wonder whether this cancelation also occurs at each higher order.
As we shall show, this is indeed the case, and one can resum a set of
diagrams analogous to Fig.~\ref{f2} but containing both resonant and
non-resonant contributions.  The result is a smoothly-varying amplitude
that agrees, to leading order in $g^2\min(\mh^2,s)/\mw^2,$ with
(\ref{e3}) in the resonant region, and (\ref{e1}) both above and below
it.

We begin by deriving this amplitude for $\VV\to\VV,$ then show how to
generalize it to full electroweak calculations of the process
$\qq\to\qq\VV$[\ref{r6}].  As a by-product, we also show how an
$s$-channel calculation can be modified to obey unitarity and more
closely reproduce the full result.  We also discuss the $\GG\to\VV$
channel and show that the same result applies there.  Finally we show
numerical results and make some concluding remarks.

\begin{figure}[b]
  \vspace{4cm}
  \centerline{
    \includegraphics{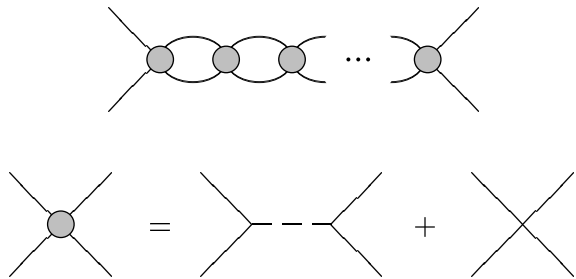}}
  \caption{All orders resummation leading to the vector boson pair
    self-energy.  Note that both resonant and non-resonant diagrams are
    included.}
  \label{f3}
\end{figure}

For $s\sim\mh^2,$ the amplitudes to scatter $\WW$ or $\ZZ$ to $\WW$ or
$\ZZ$ are all equal to the $\WW\to\ZZ$ amplitude,
$$
  i\A = \frac{-ig^2\mh^2}{4\mw^2} \frac{s}{s-\mh^2},
$$
where we have included the resonant and non-resonant contributions
without keeping track of which is which.  The full amplitude for
$\WW\to\ZZ$ is then given by a resummation analogous to Fig.~\ref{f2},
but with both resonant and non-resonant graphs appearing in each cell,
as shown in Fig.~\ref{f3}.  The result is then
$$
  i\Ab = \sum_{n=0}^{\infty}\left( \frac32 \frac1{16\pi}
    \int_{-s}^0\frac{dt}{s}i\A \right)^n i\A,
$$
where the integral is the momentum flowing around each loop, and the
factor $\frac32$ comes from the sum of $WW$ and $ZZ$ in each loop with a
factor of $\frac12$ for $ZZ$ because they are identical.  Using the
expression for the Higgs boson width,
$$
  \gh = \frac32 \frac1{16\pi} \frac{g^2\mh^3}{4\mw^2},
$$
we obtain
\beqn
  i\Ab &=& \sum_{n=0}^{\infty}
    \left( -i\frac{\gh}{\mh} \frac{s}{s-\mh^2} \right)^n
    \left( \frac{-ig^2\mh^2}{4\mw^2} \frac{s}{s-\mh^2} \right)
\nonumber\\
  &=& \frac{-ig^2\mh^2}{4\mw^2} \frac{s}{s-\mh^2+i\gh s/\mh}
\label{e4}\\
  &\equiv& \frac{-ig^2\mh^2}{4\mw^2} \frac{s}{s-\mh^2+i\Im\pvv(s)},
\nonumber
\eeqn
where we dub the function $\pvv(s)$ the vector boson pair self-energy.
Since we have
$$
  \Im \pvv(s) = \frac{\mh^2}{s}\Im \ph(s)
$$
it is clear that $\Im\pvv(\mh^2)=\Im\ph(\mh^2),$ and (\ref{e4}) is
identical to (\ref{e3}) on the resonance.

Equation (\ref{e4}) is the central result of this paper.  For
$s\gg\mh^2,$ it becomes
$$
  i\Ab \;\;{\stackrel{s\gg\mh^2}{\longrightarrow}}\;\;
\frac{-ig^2\mh^2}{4\mw^2}
    \frac1{1+i\frac32\frac1{16\pi}\frac{g^2\mh^2}{4\mw^2}}
    = \frac{-ig^2\mh^2}{4\mw^2}
    \left(1+\O\left(\frac{g^2\mh^2}{\mw^2}\right)\right),
$$
and for $s\ll\mh^2,$ it becomes
$$
  i\Ab \;\;{\stackrel{s\ll\mh^2}{\longrightarrow}}\;\; \frac{-ig^2s}{4\mw^2}
    \frac1{-1+i\frac32\frac1{16\pi}\frac{g^2s}{4\mw^2}}
    = \frac{ig^2s}{4\mw^2}
    \left(1+\O\left(\frac{g^2s}{\mw^2}\right)\right).
$$
Thus (\ref{e4}) agrees with (\ref{e1}) to leading order in
$g^2\min(\mh^2,s)/\mw^2$ above and below the resonance, and (\ref{e3})
on it, smoothly extrapolating the three regions.  We therefore describe
it as the {\em full\/} leading order amplitude for $\WW\to\ZZ$.

The amplitudes for other scattering processes $\VV\to\VV$ can be read
off from the SO(3) symmetry of the effective theory[\ref{e4}],
\beqns
  \A(\WW\to\ZZ) &\equiv& \A(s,t,u), \\
  \A(\ZZ\to\WW) &  =   & \A(s,t,u), \\
  \A(\WW\to\WW) &  =   & \A(s,t,u)+\A(t,s,u), \\
  \A(\ZZ\to\ZZ) &  =   & \A(s,t,u)+\A(t,s,u)+\A(u,t,s), \\
  \A(\ww\to\ww) &  =   & \A(t,s,u)+\A(u,t,s), \\
  \A(\WZ\to\WZ) &  =   & \A(t,s,u).
\eeqns
It is important to realize however, that
$$
  \Im\pvv(s)=0, \quad s<0,
$$
since a space-like pair cannot appear as on-shell lines in a bubble.

To translate (\ref{e4}) to the full electroweak theory, we rewrite it
\beq\label{e5}
  \frac{s}{s-\mh^2+i\gh s/\mh} =
    \frac{s^2/\mh(1+i\gh/\mh)}{s-\mh^2+i\gh s/\mh} - \frac{s}{\mh^2}.
\eeq
The apparently higher order term in the numerator is essential for the
high energy limit, and cannot be neglected.  Equation (\ref{e5})
provides a calculational implementation of (\ref{e4}) that is equally
valid in the full electroweak theory.  Namely that one makes the
replacement
$$
  \frac{i}{s-\mh^2} \to
    \frac{i(1+i\gh/\mh)}{s-\mh^2+i\gh s/\mh}
$$
for the $s$-channel Higgs boson propagator, leaving all other amplitudes
unchanged.  It would be extremely simple to make this substitution in
computer programs that calculate the {\em amplitude} for $\qq\to\qq\VV$
such as [\ref{r6}] and, with slightly more effort, in those that
directly calculate the differential cross-section.

\begin{figure}
  \centerline{
    \setlength{\unitlength}{1cm}
    \hspace{\fill}
    \begin{picture}(9,5.8)
       \put(4.5,0){\includegraphics{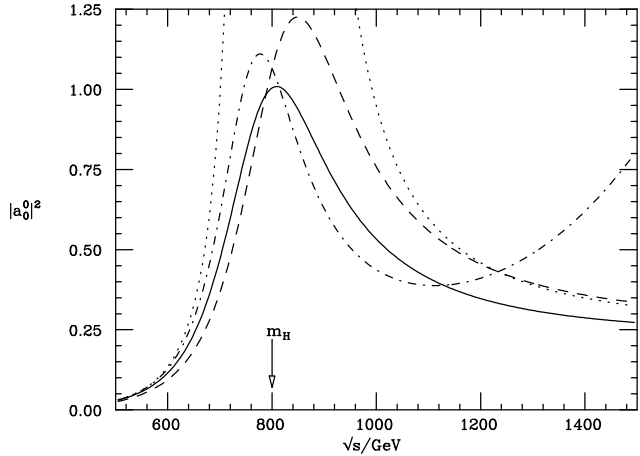}}
    \end{picture}
    \hspace{\fill}}
  \caption{The $I\!=\!0,\;J\!=\!0$ partial wave for elastic vector boson
    scattering with various treatments of the Higgs boson width: zero
    width (dotted), fixed width (dashed), using the Higgs boson
    self-energy (dot-dashed), and the full result using the vector boson
    pair self-energy (solid).}
  \label{f4}
\end{figure}

Unitarity requires that each partial wave of definite angular momentum
and isospin, $\aij,$ obeys
$$
  |\aij| \le 1.
$$
Since the condition applies to the exact amplitude, one expects small
violations at any given order in perturbation theory, owing to the
truncation of the series.  However, gross violations should be taken as
an indication of the failure of the perturbation series.  The $I\!=\!0$
case is obtained by scattering the state $(2\WW+\ZZ)/\surd6$ to itself,
and $J\!=\!0$ from the integral
$$
  \ai_0 = \frac1{16\pi}\int_{-s}^0\frac{dt}{s}\Ai.
$$
For $a^0_0,$ the only partial wave to which the Higgs resonance
contributes, we obtain
$$
  a^0_0 = -\frac{\gh}{\mh} \frac{s}{s-\mh^2+i\gh s/\mh} -
    \frac23\frac{\gh}{\mh}\left(1 -
    \frac{\mh^2}{s}\log\left(1+\frac{s}{\mh^2}\right)\right).
$$
This is shown in Fig.~\ref{f4}, in comparison with various amplitudes
that have been used in the past.  Note that only the full amplitude
satisfies unitarity both in the resonance region and well above it.
Note also that it peaks very close to $\mh,$ unlike the other cases.

In the resonance region all the possibilities (except the divergent one)
are equally valid at leading order, but show marked differences in the
lineshape, indicating the need for a full next-to-leading order
calculation.  However, since the full amplitude is correct above, below
and on the resonance, we expect it to give the most accurate lineshape.

\begin{figure}
  \centerline{
    \setlength{\unitlength}{1cm}
    \hspace{\fill}
    \begin{picture}(9,5.8)
       \put(4.5,0){\includegraphics{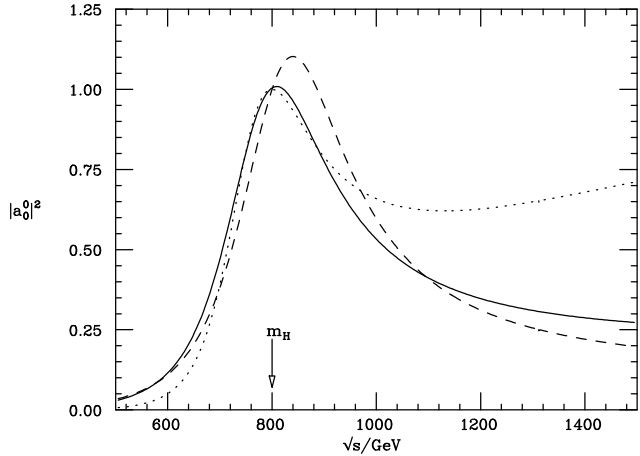}}
    \end{picture}
    \hspace{\fill}}
  \caption{As in Fig.~\ref{f4}, using the full result (solid), the
    na\"\i ve $s$-channel approximation (dotted) and the improved
    $s$-channel approximation (dashed).}
  \label{f5}
\end{figure}

Comparing equations (\ref{e4}) and (\ref{e5}), we see the opportunity to
make an improvement to the $s$-channel approximation.  The $s$-channel
approximation consists of using {\em only} the diagrams in which the
$s$-channel Higgs boson propagator appears, i.e.~it gives us the first
term of (\ref{e5}).  If we multiply this by $\mh^2/s$ instead of
$(1+i\gh/\mh),$ we obtain exactly (\ref{e4}).  Thus in the $\WW\to\ZZ$
case, this improved $s$-channel approximation is exact.  We show
numerical results for the $I\!=\!0$ case in Fig.~\ref{f5}.

\begin{figure}[b]
  \vspace{2cm}
  \centerline{
    \includegraphics{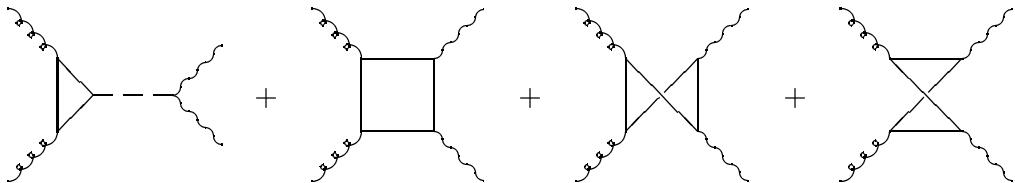}}
  \caption{Lowest order Feynman diagrams for $\GG\to\ZZ$.}
  \label{f6}
\end{figure}

We turn now to the gluon fusion process.  Although the coupling of
gluons to electroweak bosons, which is mediated by quark loops, is
rather weak, the high density of gluons within a hadron means that this
is a competitive source of vector boson pairs.  The lowest order
diagrams are shown in Fig.~\ref{f6}.  The amplitude is[\ref{r7}]
\beqns
  i\B &=& \frac{-ig^2g_s^2}{\mw^2} \left\{\frac{s^2}{s-\mh^2}-s\right\}
    I(s), \\
  I(s) &=& \frac12\frac{m_q^2}{s}
    \left[ \left(\log\frac{s}{m_q^2}-i\pi\right)^2 -4\right].
\eeqns
Since this has the identical form to (\ref{e1}), it is clear that
exactly the same conclusions will apply:  the resonant and non-resonant
diagrams can be canceled at high energy in each order;  they can be
resummed to all orders;  the result can be implemented by the
replacement
$$
  \frac{i}{s-\mh^2} \to
    \frac{i(1+i\gh/\mh)}{s-\mh^2+i\Im\pvv(s)}.
$$
Since gluons and vector bosons couple together so weakly, we neglect the
effect of internal gluon lines, so
$$
  \Im\pvv(s) = \gh s/\mh
$$
exactly as before.

\begin{figure}
  \centerline{
    \setlength{\unitlength}{1cm}
    \hspace{\fill}
    \begin{picture}(9,5.8)
      \put(4.5,0){\includegraphics{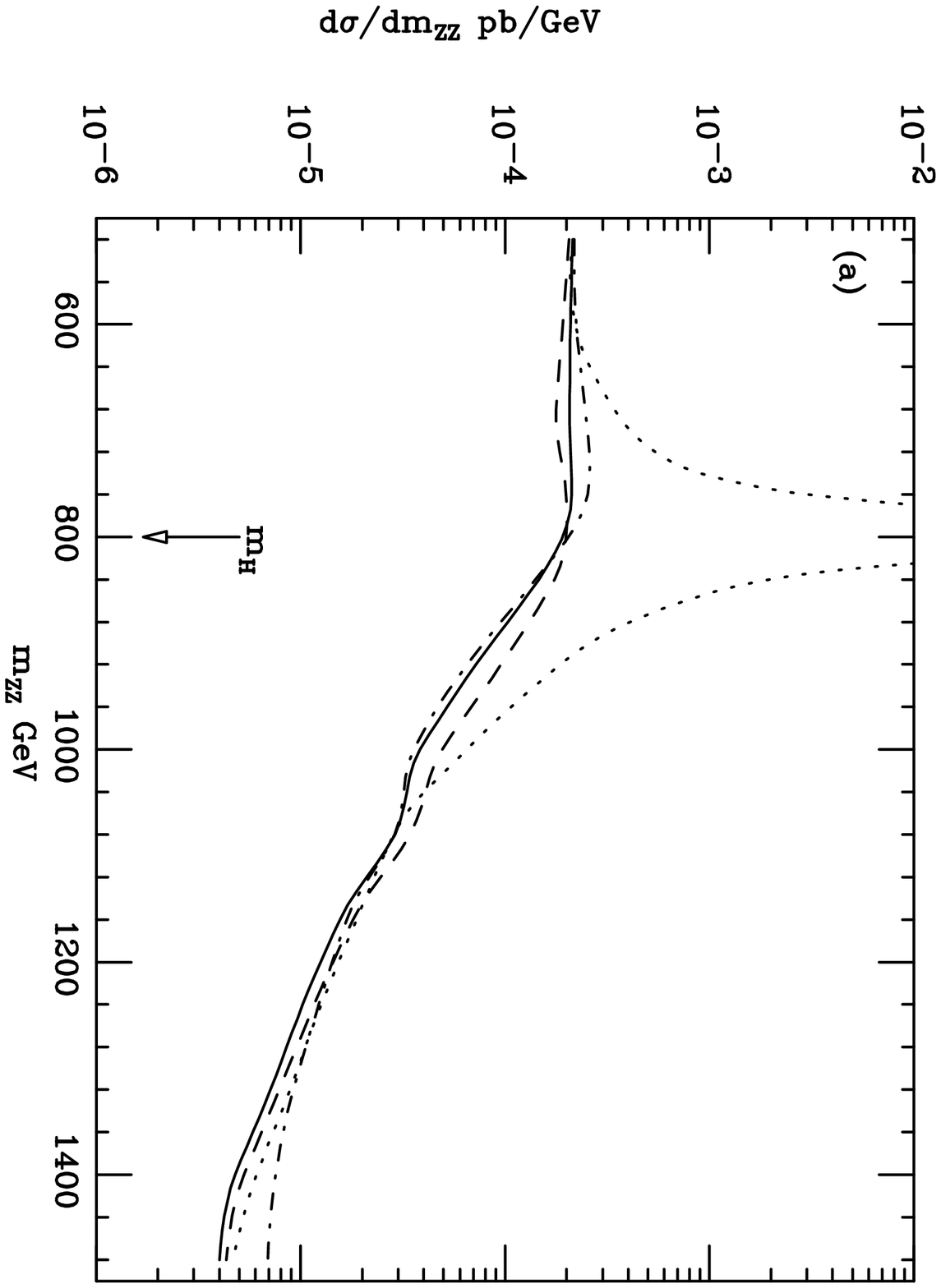}}
    \end{picture}
    \hspace{\fill}\hspace{\fill}
    \begin{picture}(9,5.8)
       \put(4.5,0){\includegraphics{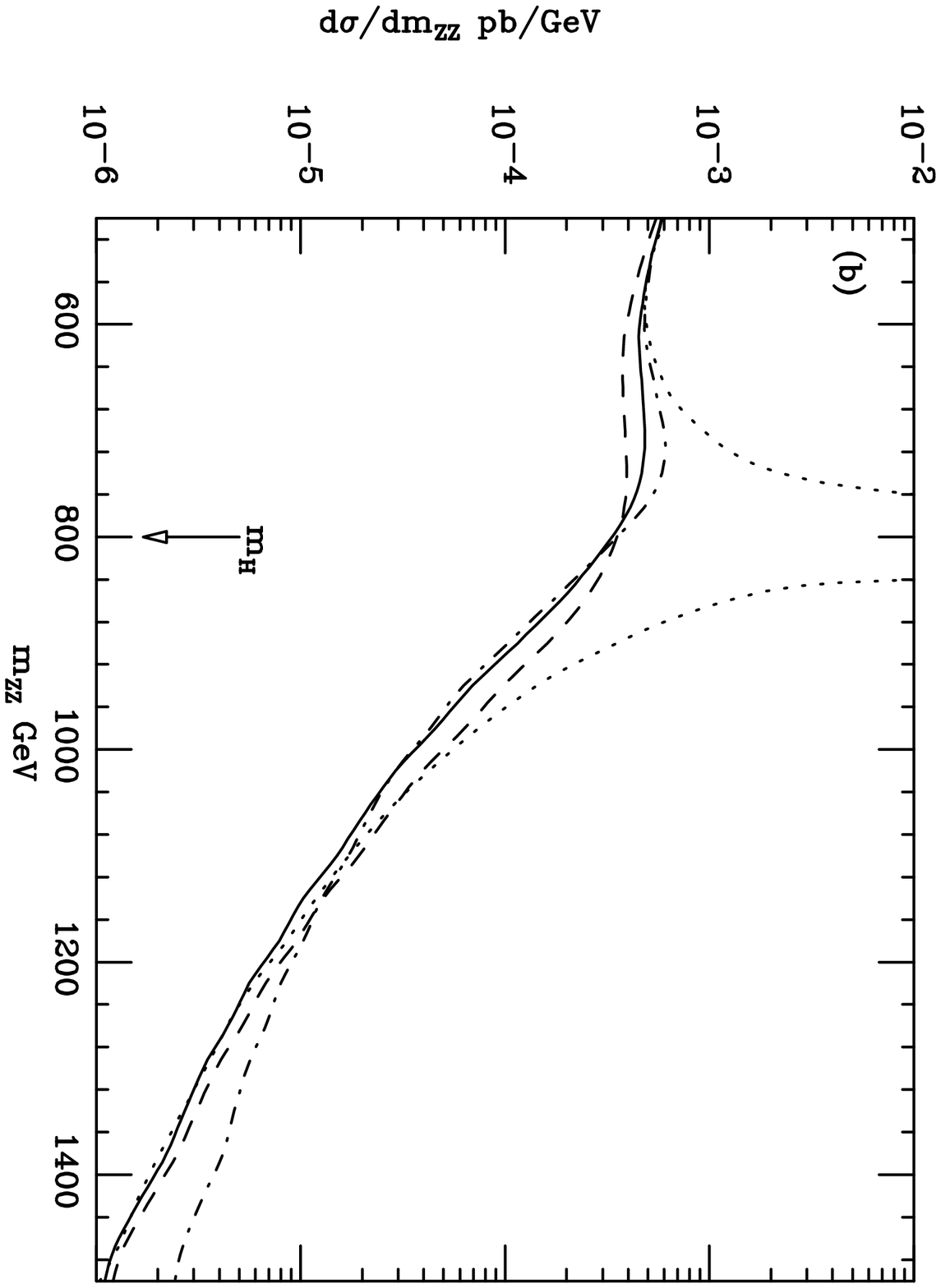}}
    \end{picture}
    \hspace{\fill}}
  \caption{The $\ZZ$ invariant mass spectrum at the LHC from (a)
    $\qq\to\qq\ZZ$ and (b) $\GG\to\ZZ$.  We set $m_t=175$~GeV,
    $\mz=91.2$~GeV, $\alpha=1/128,$ $\sin^2\theta_w=0.23,$
    $\mw=\mz\cos\theta_w$ and $\alpha_s(\mz)=0.120,$ and use the MRS
    D--$'$ parton distribution functions.  Curves are as in
    Fig.~\ref{f4}.}
  \label{f7}
\end{figure}

We have modified the programs of~[\ref{r6}] for $\qq\to\qq\VV$ and
[\ref{r8}] for $\GG\to\VV$ according to this prescription, and the
results are shown for the LHC in Fig.~\ref{f7}.  It can clearly be seen
that the differences in lineshape and behaviour well above the resonance
persist even in the full electroweak calculations convoluted with parton
densities.  Owing to the fall of parton densities with increasing
energy, the full result no longer peaks at the Higgs mass.

\begin{figure}
  \centerline{
    \setlength{\unitlength}{1cm}
    \hspace{\fill}
    \begin{picture}(9,5.8)
       \put(4.5,0){\includegraphics{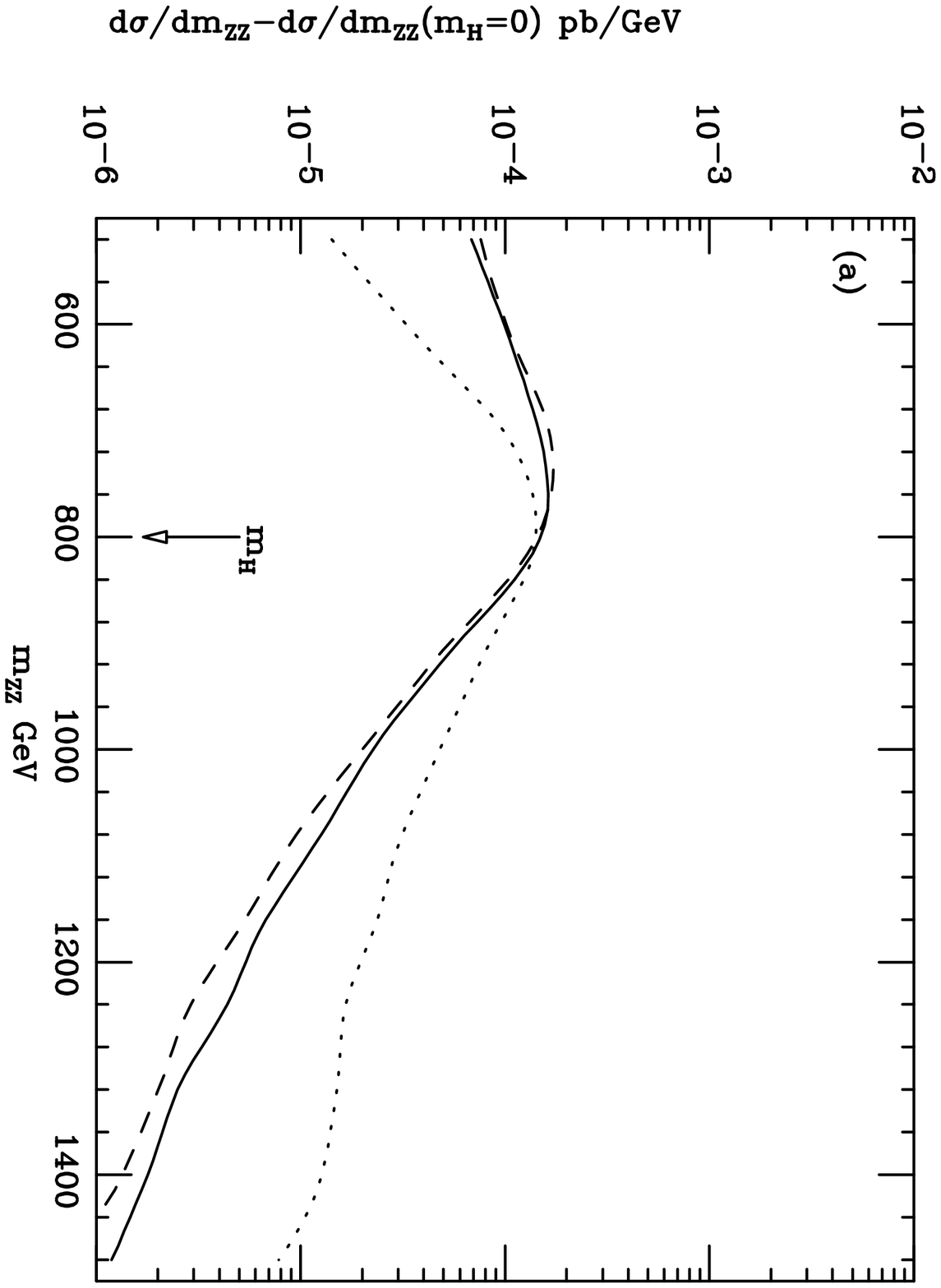}}
    \end{picture}
    \hspace{\fill}\hspace{\fill}
    \begin{picture}(9,5.8)
       \put(4.5,0){\includegraphics{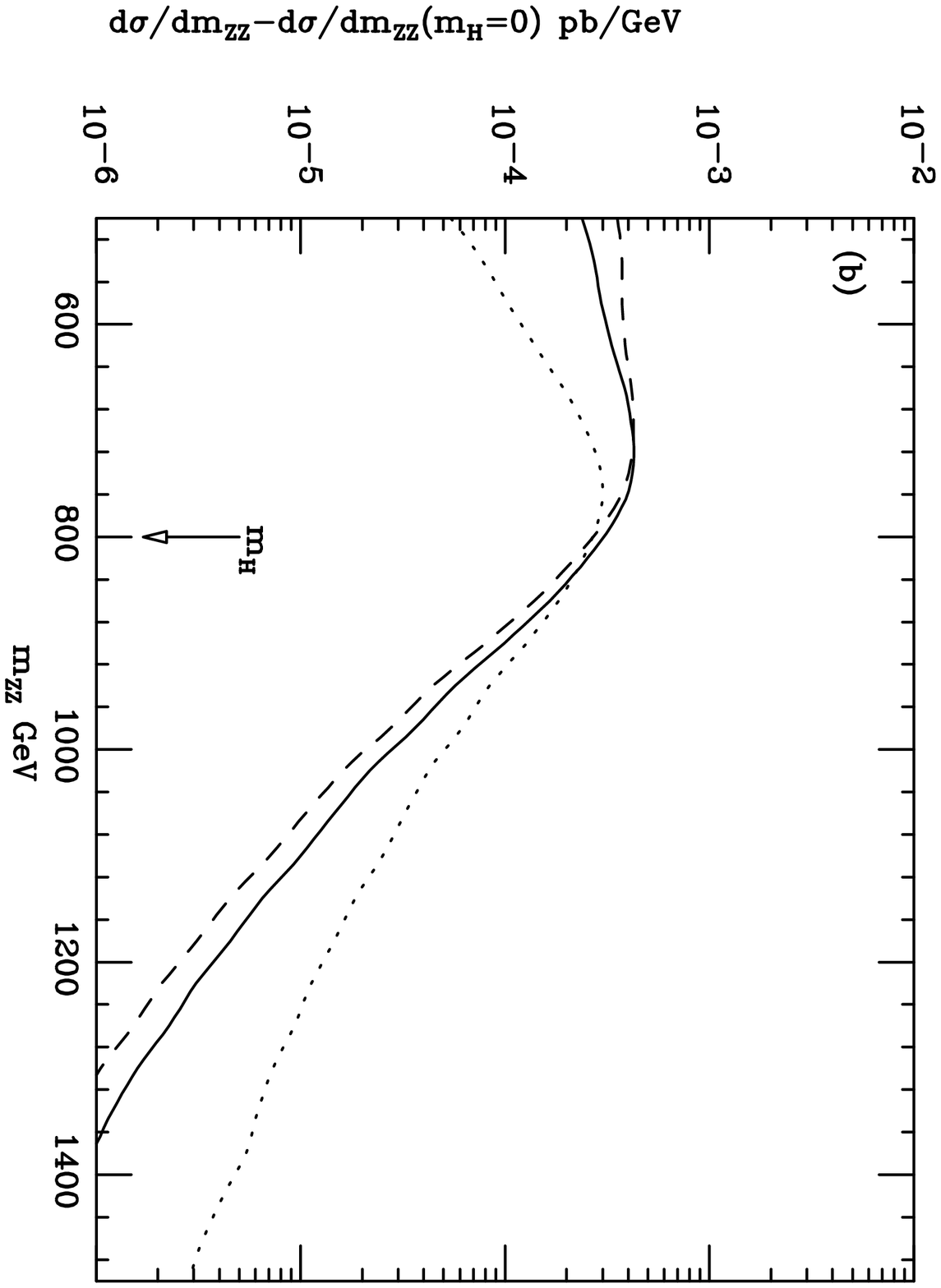}}
    \end{picture}
    \hspace{\fill}}
  \caption{The excess of the $\ZZ$ invariant mass spectrum over the
    $\mh\to0$ expectation at the LHC from (a) $\qq\to\qq\ZZ$ and (b)
    $\GG\to\ZZ$.  Curves are as in Fig.~\ref{f5} and parameters as in
    Fig.~\ref{f7}.}
  \label{f8}
\end{figure}

We would also like to compare the full result with our improved
$s$-channel approximation.  However, since the $s$-channel approximation
is only intended to model the Higgs boson `signal', and not the
$\O(g^2)$ `background' we compare it with the full result after
subtraction of this background.  As usual[\ref{r9}], we define the
background to be the full result in the limit $\mh\to0,$ as this gives
the lowest rate one could expect.  It is clear from (\ref{e1}) that this
background is zero in the effective theory.  The comparison is shown in
Fig.~\ref{f8}, where it can be seen that the improved $s$-channel
approximation performs much better than the na\"\i ve one.

To conclude, the principal result of this paper is shown in
Fig.~\ref{f3} and Eq.~(\ref{e4}).  It is that it is possible to resum
the sum of resonant and non-resonant diagrams to all orders, and the
result smoothly extrapolates the well-known correct behaviour below,
above and on the resonance.  As a calculational prescription, it is
possible to represent the result as a modification of the Higgs boson
propagator,
$$
  \frac{i}{s-\mh^2} \to
    \frac{i(1+i\gh/\mh)}{s-\mh^2+i\gh s/\mh},
$$
although it should be stressed that it includes effects that are not
strictly associated with the propagation of a Higgs boson, namely the
interference with non-resonant diagrams.  In calculations that use the
$s$-channel approximation, a better modification is
$$
  \frac{i}{s-\mh^2} \to
    \frac{i\mh^2/s}{s-\mh^2+i\gh s/\mh}.
$$
We have shown that the impact on the Higgs boson lineshape, and hence on
the whole phenomenology of high energy vector boson pair production, is
significant.

\subsection*{Acknowledgements}
I am grateful to Nigel Glover for many useful discussions, and for
providing the computer programs used for Figs.~\ref{f7} and~\ref{f8}.

\subsection*{References}
\begin{enumerate}
\item\label{r1}
  See for example, J.F.~Gunion et al., {\em The Higgs Hunters' Guide}
  (Addison Wesley, 1990), and references therein
\item\label{r2}
  M.S.~Chanowitz and M.K.~Gaillard, Nucl.~Phys.~B261 (1985) 379
\item\label{r3}
  G.~Valencia and S.S.D.~Willenbrock, Phys.~Rev.~D42 (1990) 853
\item\label{r4}
  S.~Dawson and S.S.D.~Willenbrock, Phys.~Rev.~D40 (1989) 2880
\item\label{r5}
  G.~Valencia and S.S.D.~Willenbrock, Phys.~Rev.~D46 (1992) 2247
\item\label{r6}
  U.~Baur and E.W.N.~Glover, Nucl.~Phys.~B347 (1990) 12
\item\label{r7}
  E.W.N.~Glover and J.J.~van der Bij, Phys.~Lett.~B219 (1989) 488
\item\label{r8}
  E.W.N.~Glover and J.J.~van der Bij, Nucl.~Phys.~B321 (1989) 561
\item\label{r9}
  E.W.N.~Glover, in {\em Proc.~26$^{th}$ Rencontre de Moriond, High
    Energy Hadronic Interactions}, ed.~J.~Tran Thanh Van (Editions
  Fronti\`eres, 1991), p.~161
\end{enumerate}

\end{document}